\documentclass[aps,jmp,amsmath,amssymb,reprint]{revtex4-1}
\usepackage{graphicx}
\usepackage{dcolumn}
\usepackage{bm}
\usepackage{graphicx}
\usepackage{subfigure}
\usepackage{physics}
\usepackage[usenames, dvipsnames]{color}
\usepackage{hyperref}
\usepackage[utf8]{inputenc}

\begin{document}

\title{Carrier localization and miniband modeling of InAs/GaSb based type-II superlattice infrared detectors}          

	\author{Swarnadip Mukherjee}
	\thanks{corresponding author: swarnadip@ee.iitb.ac.in}
	\author{Anuja Singh}
	\author{Aditi Bodhankar}
	\author{Bhaskaran Muralidharan}
	\affiliation{Department of Electrical Engineering, Indian Institute of Technology Bombay, Powai, Mumbai-400076, India}
\date{\today}

\begin{abstract}
Microscopic features of carrier localization, minibands, and spectral currents of InAs/GaSb based type-II superlattice (T2SL) mid-infrared detector structures are studied and investigated in detail. In the presence of momentum and phase-relaxed elastic scattering processes, we show that a self-consistent non-equilibrium Green's function method within the effective mass approximation can be an effective tool to fairly predict the miniband and spectral transport properties and their dependence on the design parameters such as layer thickness, superlattice periods, temperature, and built-in potential. To benchmark this model, we first evaluate the band properties of an infinite T2SL with periodic boundary conditions, employing the envelope function approximation with a finite-difference discretization within the perturbative eight-band $\bf{k.p}$ framework. The strong dependence of the constituent material layer thicknesses on the band-edge positions and effective masses offers a primary guideline to design performance-specific detectors for a wide range of operation. Moving forward, we demonstrate that using a finite T2SL structure in the Green's function framework, one can estimate the bandgap, band-offsets, density of states, and spatial overlap which comply well with the $\bf{k.p}$ results and the experimental data. Finally, the superiority of this method is illustrated via a reasonable estimation of the band alignments in barrier-based multi-color non-periodic complex T2SL structures. This study, therefore, provides deep physical insights into the carrier confinements in broken-gap heterostructures and sets a perfect stage to perform transport calculations in a full-quantum picture. 
\end{abstract}

\maketitle

\section{Introduction}
Type-II superlattices (T2SL) made of InAs/(In,Ga)Sb \cite{SaiHalasz} have found immense applications in the field of infrared (IR) detectors \cite{Smith_t2sl,razeghiapl2005} and focal plane arrays (FPA) \cite{256FPA,gunapalaFPA} for defense, space, and biomedical applications over the conventional mercury cadmium telluride (MCT) \cite{RogalskiHgCdTe}, quantum-well (QW) \cite{QWIP_levine}, and quantum dot (QD) \cite{QDMARTYNIUK,LIUQDIP} technologies due of their fine band tuning, flexible band alignment, moderate dark current, suppressed Auger recombination, high quantum efficiency and established III-V growth technology \cite{Martyniuk_review,Rogalski_review,plis2014inas}. The excellent band-tunability  features of highly lattice-matched $6.1${\AA} family of III-V compound semiconductors \cite{KROEMER6.1} enabled in covering a wide range of infrared spectrum spanning from mid-infrared $3-5$ $\mu m$ (MWIR) to long-infrared $8-15$ $\mu m$ (LWIR). The research in this field has evolved rapidly over the last few years and has taken a giant leap after the discovery of barrier-based (nBn \cite{nBn,nBn_rodriguez}, pBp \cite{pBp}, XBn \cite{XBn}, CBIRD \cite{cbird}, pBiBn \cite{pbibn} etc.), complex (M \cite{M}, N \cite{N}, W \cite{W} structures) and cascaded SL \cite{cascade} architectures which made detectors to operate at higher operating temperatures and at higher efficiencies.\\ 
\indent The key factors in designing detectors using quantum confined SL structures for a specific range of wavelength are the right choice of the constituent materials, their arrangements and tuning of the growth parameters to have a precise control over their thicknesses. Moreover, in T2SL, a special care must be taken owing to their broken-gap band alignment to achieve adequate absorption. The band properties exhibited by T2SL is primarily dependent on the design parameters which led several groups to study their electronic band structure using different approaches like $\bf{k.p}$ \cite{livneh2012k,klipstein2010operator,klipstein2013k,qiao2012electronic}, empirical tight-binding \cite{razeghi_tb,razeghi2010band}, pseudopotential \cite{pseudopotential} and density functional theory \cite{garwood2017electronic}. However, consolidated studies on the variation of the key band parameters regulated by the layer thicknesses for a wide range of operation are still very scarce in the literature.\\
\indent In this work, the $\bf{k.p}$ approach with envelope function approximation (EFA) is chosen over other methods to calculate the electronic structure due to its higher accuracy in predicting SL bandgap and excellent handling of interface \cite{HauganPRB} to balance the strain effects. The discretization of the envelope functions along the out-of-plane direction is incorporated though the finite difference method (FDM) \cite{FDM,klipstein2013k,akhavan_effectivemass} for its simplicity and accuracy around the high-symmetry point over other techniques like Fourier transform method (FTM) \cite{FTM} and finite element method (FEM) \cite{becer2019modeling}, etc. A proper treatment of the interface is taken by adopting the operator-ordered interface matrix \cite{klipstein2010operator,klipstein2013k} approach along with the inclusion of a separate interfacial layer within the unit cell. From the design point of view, our results primarily put emphasis on the variations of the miniband positions and the density of states (DOS) effective masses with respect to the layer thicknesses. While the bandgap and miniband widths determine the spectral range of operation, the DOS effective mass plays a vital role in determining the quantum efficiency through its strong dependence on the diffusion length and the quantum mechanical tunneling probability of carriers that directly affects the transport properties.\\
\indent An in-depth understanding of the microscopic picture of the quantum kinetic properties of nanoscale materials and devices requires a pure quantum mechanical modeling of carrier transport. A rapid advancement in research along this direction has been experienced since the emergence of the non-equilibrium Green's function (NEGF) based ballistic as well as diffusive quantum transport models, widely implemented in several applications like spintronics \cite{abhishek_prapp,debasish}, thermoelectrics \cite{myTED,myPRA,myTF,pankaj,Aniket2017}, optoelectronics \cite{Henrickson,Cavassilas_NEGF,Aeberhard_jce,PRL_nanotube}, memory devices \cite{jyotsna}, to name a few. Specifically, it has received a lot of recent attention in demonstrating the carrier localization and non-equilibrium transport properties of finite-sized superlattice based solar cells \cite{AeberhardPRB2008,Aeberhard_jce}, photodetectors \cite{akhavanted} and quantum cascade lasers \cite{QCLnegf}.\\
\indent With this motivation, we employ a NEGF based quantum transport model withing the single-band effective mass approximation \cite{akhavan_effectivemass,Kolek_effmass,DattaQT} to investigate the miniband and spectral transport properties \cite{Kaya_miniband,miniband_spie,Aeberhard_apl} of a finite T2SL structure. In the presence of momentum and phase-breaking elastic scattering processes \cite{DattaQT,dattaLNE,abhishek_ted,abhishek_prapp,abhishek_apl}, the NEGF results evidence a close agreement with the $\bf{k.p}$ results for an infinite SL. A systematic study on the DOS, carrier localization and spectral current is also presented for a qualitative understanding of the non-trivial roles played by the thickness parameters, built-in potential and temperature in tailoring the spatial separation of carrier confinement and coupling of periods. The applicability of this method is further shown to extract adequate information on the band alignment properties of the multi-band unipolar or bipolar barrier-based complex T2SL structures of varying SL configuration with broken periodicity. Finally, employing a simple interface modeling technique, we demonstrate that the bandgap can also be predicted using the NEGF method within an acceptable range which was overestimated earlier in the absence of interface modeling.\\ 
\indent The rest of the paper is organized as follows. In Sec. \ref{sec_theory}, we briefly outline the $\bf{k.p}$ and NEGF based theoretical models respectively used to calculate the band structure and miniband characteristics. We discuss the results in Sec. \ref{sec_result} which is divided into two subsections, namely, Sec. \ref{sec_kp} for $\bf{k.p}$ results and Sec. \ref{sec_negf} for NEGF results. We conclude this paper in Sec. \ref{conclu}.
\section{Theory and Model}
\label{sec_theory}
\subsection{k.p Method}
The $\bf{k.p}$ perturbation theory is one of the widely studied methods for implementing the electronic band structure of the III-V and II-VI group of compound materials and their hetero-structures. The supremacy of this method in computing bandgap and oscillator strengths has proven its immense capabilities in studying the optoelectronic devices \cite{klipstein2010operator,klipstein2013k,livneh2012k,qiao2012electronic}. Extensive research has been carried out in understanding the oscillator strengths in order to compute the absorption rate of T2SL based infrared photodetectors \cite{livneh2012k,smith1990theory,qiao2012electronic}. In this paper, a comprehensive formulation of the $\bf{k.p}$ Hamiltonian of InAs/GaSb superlattice has been presented.\\
\begin{figure}[!tbp]
	\centering
	\includegraphics[height=0.2\textwidth,width=0.35\textwidth]{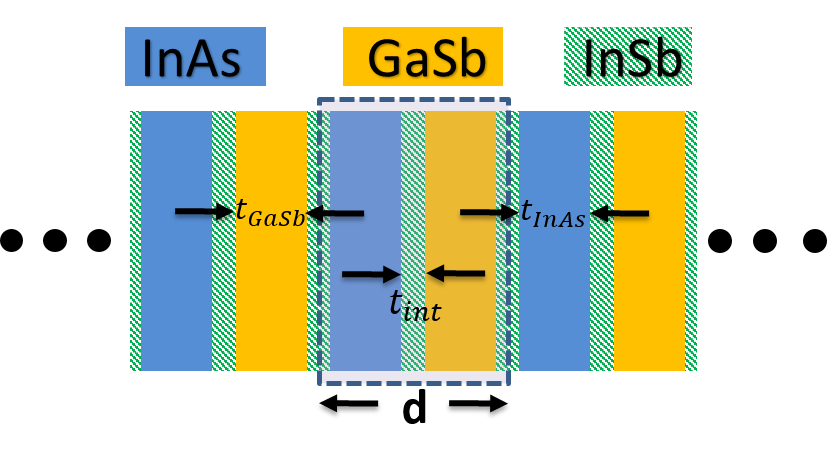}
	\caption{T2SL schematic: InAs/GaSb based T2SL structure with InSb type interface. The unit cell of thickness $d=t_{InAs}+t_{GaSb}$ consists of InAs, InSb and GaSb layers with effective thicknesses of $t_{InAs}-t_{int}$, $2t_{int}$, and $t_{GaSb}-t_{int}$, respectively, arranged as shown.}
	\label{uc}
\end{figure}
\indent Initially, a Hamiltonian for the bulk material is implemented based on Kane's model and its extension to the Lowdin's Perturbation theory for degenerate bands \cite{kane1980band}. Spin-orbit interactions are also taken into account for the triply degenerate valence bands. The interpolation of the same method to complex hetero-structures like superlattices, quantum wells etc. is done using the EFA technique \cite{bastard1981superlattice}. Within the EFA theory, the superlattice wavefunctions ($\Psi_{n}(z)$) are expressed in terms of slowly varying envelope functions ($F(z)$) along the growth direction (here, z-direction) in the orbital basis states ($u_0(z)$) \cite{bastard1981superlattice} and are given by
\begin{equation}
	\Psi_{n}(z) = \sum_{j}^{}F_{j}(z)u_{j0}(z).
	\label{eq1}	
\end{equation}
These envelope functions are subject to periodic boundary conditions. For a given SL with $N_{z}$ grid points having a period thickness of $d$, one can write
\begin{equation}
	\begin{split}
		&F_{j}^{i}(k_{t},z_{0}) = e^{-iqd}F_{j}^{i}(k_{t},z_{N_{z}}),\\
		&F_{j}^{i}(k_{t},z_{N_z+1}) = e^{iqd}F_{j}^{i}(k_{t},z_{1})
	\end{split}
	\label{2}
\end{equation}
where $k_t$ is the transverse momentum and $q$ is the Bloch vector of $F(z)$ that spans the Brillouin zone (BZ). The spatial dependence of the material properties results in recognizing the quantum number $k_z$ as differential operators rather than constants which necessitates the implementation of an operator ordering technique in order to maintain the hermiticity of the overall Hamiltonian \cite{burt1992justification}. One of the most widely studied operator ordering methods uses the symmetrization technique which does not accurately estimate the inter-band coupling in the case of heavy hole bands unlike Burt-Foreman's proposed method of non-symmetry \cite{foreman1993effective}. Since the material properties repeat periodically along the [001] crystalline plane, the quantum number $k_{z}$ is modified to a differential operator as $-i\frac{\partial}{\partial z}$. The resultant Hamiltonian of the SL consisting of differential terms up to 2\textsuperscript{nd} order, is expressed as a sum of three matrices ($H^{(0)}$, $H^{(1)}$ and $H^{(2)}$), given by $H(k_t,k_z)=H^{(0)}+H^{(1)}\left(-i\frac{\partial}{\partial z}\right)+\left(-i\frac{\partial}{\partial z}\right)H^{(2)}\left(-i\frac{\partial}{\partial z}\right)$. Hence, a finite difference discretization technique \cite{FDM} is employed here in converting the coupled differential equations into a set of algebraic equations that can be solved numerically to find the envelope functions.\\
\indent In semiconductor superlattices having abrupt and non-symmetric interfaces, the interfacial region plays a crucial role in obtaining accurate bandgaps \cite{haugan2004band}. To model the interface, a thin layer of InSb is intentionally inserted (Fig. \ref{uc}) at the junction of InAs and GaSb in order to compensate the strain effects arising due to the lattice mismatched InAs layers on GaSb substrate. Hence, in the formulation presented in this paper, both operator ordering and interface Hamiltonian ($H_{IF}$), given by
\begin{equation}
	H_{IF}=\sum_{z_i}\delta(z-z_i)\begin{bmatrix}
		D_S & 0 & 0 & \pi_i \beta \\ 0 & D_X & \pi_i \alpha & 0 \\
		0 & \pi_i \alpha & D_X & 0 \\ \pi_i \beta & 0 & 0 & D_Z
	\end{bmatrix},
\end{equation}
have been considered \cite{klipstein2010operator,klipstein2013k,delmas2019comprehensive}. Here, the interface parameters $\alpha$ and $\beta$ have been set to 0.2 $eV${\AA} and the diagonal fitting parameters $Ds$, $Dx$, $Dz$ are respectively taken as 3 $eV${\AA}, 1.3 $eV${\AA} and 1.1$eV${\AA} \cite{livneh2012k}. The original $\bf{k.p}$ Hamiltonian $H(k_t,k_z)$ considered in this work, is formed in terms of the standard $F$, $G$, $H$ parameters \cite{Galeriu2005}. 
\subsection{NEGF Method}
The quantum mechanical properties of a low-dimensional device connected to several macroscopic contacts and subject to different scattering phenomena can be attributed by the retarded Green's function ($G^R$) with the proper description of different self-energies ($\Sigma$). For the given SL structures as shown in Fig.~\ref{ucn}, $G^R$ along the direction of longitudinal energy ($E$) is defined as
\begin{equation}
G^R_b(z,z',E)=[E^+I-H_b-U_b-\sum_{j}\Sigma^C_{j,b}(E)-\Sigma^S_b(E)]^{-1},
\label{green}
\end{equation}
where $b$ is the band index, $z,z'$ are the position index, $E^+=E+i\eta^+$ where $\eta^+$ is a small positive number, $I$ is the Identity matrix, $H$ is the real space 1-D tight-binding Hamiltonian of the SL, $U$ is the potential profile ideally obtained from the self-consistent NEGF-Poisson solver, $\Sigma^C_j$ is the self-energy of the $j^{th}$ contact (where $j\in L$ (left), $R$ (right)) and $\Sigma^S$ is the scattering self-energy. The active SL region is modeled using the nearest neighbor tight-binding Hamiltonian ($H$) with decoupled single-band effective mass approximation. At zero transverse momentum ($k_{\parallel}=0$), the elements of the $i^{th}$ row of $H$ are expressed as $H_{i,i}=\frac{\hbar^2}{\Delta z^2}\left(  \frac{1}{m^*_{i-1}+m^*_{i}}+\frac{1}{m^*_{i+1}+m^*_{i}} \right)$, $H_{i,i+1}=\frac{\hbar^2}{\Delta z^2}\left(  \frac{1}{m^*_{i+1}+m^*_{i}} \right)$, and $H_{i,i-1}=\frac{\hbar^2}{\Delta z^2}\left(  \frac{1}{m^*_{i-1}+m^*_{i}} \right)$, where $\Delta z$ is the lattice spacing along the $z$-direction, $\hbar$ is the reduced Planck's constant and $m^*_i$ is the electron or hole effective mass of the material corresponding to $i^{th}$ atom of the 1-D SL chain. The local density of states (LDOS) at an energy $E$ is given by the diagonal elements of the spectral function, defined as
\begin{equation}
	A_b(z,z',E)=i \left[ G_b(z,z',E)-G_b^{\dagger}(z,z',E) \right].
	\label{LDOS}
\end{equation} 
The self-energies of the semi-infinite contacts for a given band are calculated from the out of plane contact dispersion relations \cite{DattaQT}. The imaginary part of $\Sigma^C$ quantifies the broadening matrix as $\Gamma^C=i(\Sigma^C-\Sigma^{C\dagger})$. In the absence of scattering, the transmission function between the two contacts for any band $b$ is given by \cite{DattaQT}
\begin{equation}
	T_b(E)=Re[Tr(\Gamma^C_{1,b}G^R_b\Gamma^C_{2,b}G^A_b)].
	\label{transeq}
\end{equation}
In the scattering picture, $\Sigma^S$ can be calculated self-consistently from the knowledge of the correlation functions of electrons ($G^n$) and holes ($G^p$). These correlation functions are  evaluated from the total in ($\Sigma^{in}$) and out-scattering ($\Sigma^{out}$) self-energies of the device and are expressed as 
\begin{equation}
	\begin{split}
		&G^n(E)=G(E)\Sigma^{in}(E)G^{\dagger}(E),\\ &G^p(E)=G(E)\Sigma^{out}(E)G^{\dagger}(E),
	\end{split}
	\label{GnGp}
\end{equation}
where $\Sigma^{in}(E)=\sum_{j=L,R}\Gamma^C_j(E)f_j(E-\mu_j)+\Sigma^{in}_S$ and $\Sigma^{out}(E)=\sum_{j=L,R}\Gamma^C_j(E)\left(1-f_j(E-\mu_j)\right)+\Sigma^{out}_S$. Here, $f_j(E-\mu_j)$ is the Fermi function of the $j^{th}$ contact having electrochemical potential $\mu_j$ and $\Sigma_S^{in(out)}$ is the in (out) scattering self-energy for the scattering event $S$. In this work, we have considered the phase and momentum-breaking elastic scattering events. Within the elastic limit of dephasing \cite{dattaLNE,abhishek_ted,abhishek_prapp,abhishek_apl}, $\Sigma_S^{in(out)}$ simply becomes proportional to $G^{n(p)}$ with a scaling factor of $D_0$, defined via the strength of scattering which in our simulation is assumed to be energy independent and is set equal to $10^{-5}$ $eV^2$. Having obtained $\Sigma_S^{in(out)}$, one can evaluate $\Sigma^S$ as $\Sigma^S(E)=i\Im[\Sigma^S(E)]\ast \left(\delta(E)+\frac{i}{\pi E}\right)$, where $\ast$ denotes convolution and $\Im[\Sigma^S(E)]=-\frac{i}{2}\left(\Sigma_S^{in}+\Sigma_S^{out}\right)$. Having obtained the correlation functions self-consistently with the scattering functions, one can express the spectral current flowing from point $z_j$ to $z_{j+1}$ as \cite{DattaQT}
\begin{equation}
	I^{sp}_{el(hh)}(E)=\frac{iq}{\pi \hbar}\left[ H_{j,j+1}G^{n(p)}_{j+1,j}(E)\\-H_{j+1,j}G^{n(p)}_{j,j+1}(E) \right].
	\label{Ieqn}
\end{equation}

\begin{figure}[!htbp]
	\centering
	\subfigure[]{\includegraphics[height=0.225\textwidth,width=0.225\textwidth]{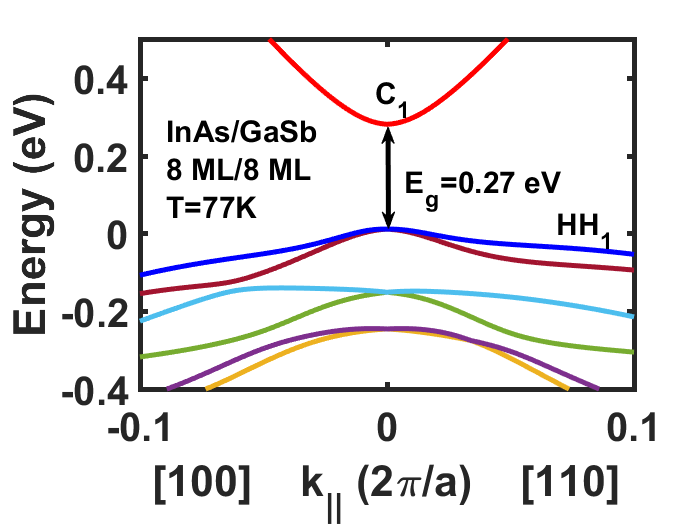}\label{Eka}}
	\quad
	\subfigure[]{\includegraphics[height=0.225\textwidth,width=0.225\textwidth]{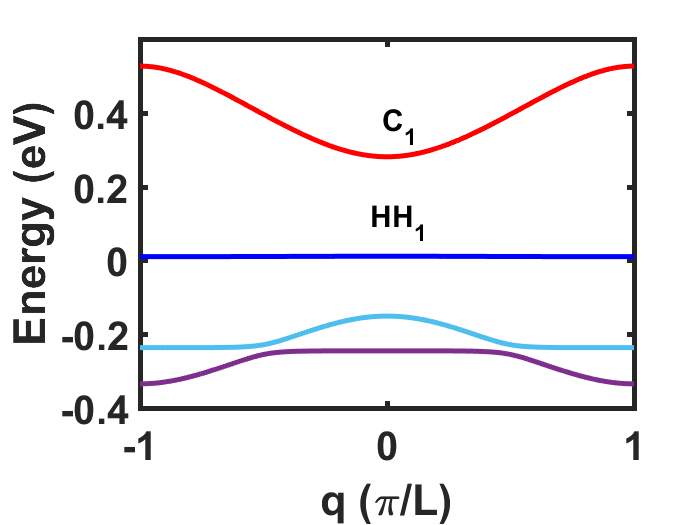}\label{Ekb}}
	\quad
	\caption{Electronic band structure of (8 ML/8 ML) InAs/GaSb based T2SL at $T=77$ K using $8\times8$ $\bf{k.p}$ technique: (a) In-plane and (b) out-of-plane dispersion in the first Brillouin zone using periodic boundary condition of the given T2SL (period=16 ML) with InSb interface of 0.5 ML. Calculated bandgap $E_g=0.27$ $eV$ perfectly matches with the measured experimental value. The band curvatures at the high symmetry $\Gamma$ point ($k=0$) indicates the low electron effective mass and very high heavy hole effective mass along the out-of-plane direction.}
	\label{Ek}
\end{figure}
\begin{figure*}[!htbp]
	\centering
	\subfigure[]{\includegraphics[height=0.25\textwidth,width=0.25\textwidth]{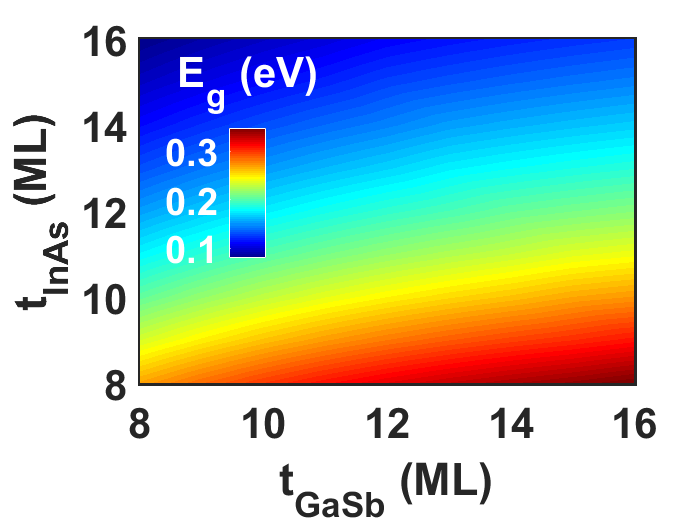}\label{Ega}}
	\quad
	\subfigure[]{\includegraphics[height=0.25\textwidth,width=0.25\textwidth]{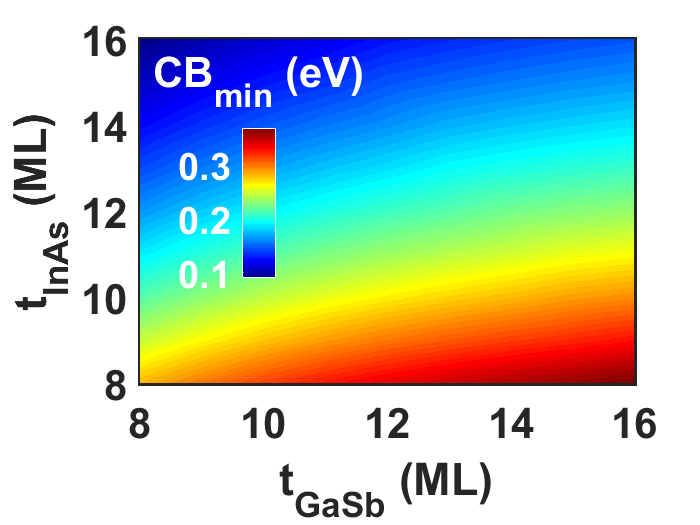}\label{Egb}}
	\quad
	\subfigure[]{\includegraphics[height=0.25\textwidth,width=0.25\textwidth]{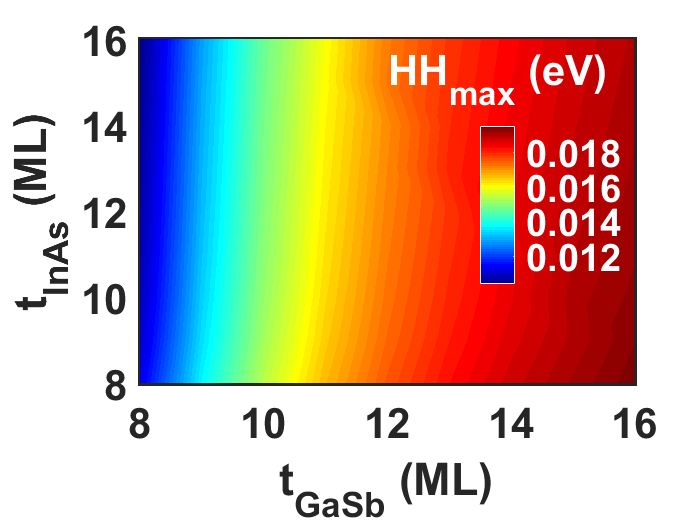}\label{Egc}}
	\quad
	\caption{T2SL bandgap and band offsets as functions of $t_{InAs}$ and $t_{GaSb}$ in 2D color plots: (a) Bandgap ($E_g$), and (b) conduction band minima ($CB_{min}$) and (c) heavy hole band maxima ($HH_{max}$) with respect to the zero energy reference are plotted as functions of $t_{InAs}$ and $t_{GaSb}$ (in ML). $E_g$ and $CB_{min}$ exhibit similar patterns and are sensitive to the variations of both $t_{InAs}$ and $t_{GaSb}$. In contrast, $HH_{max}$ varies only with $t_{GaSb}$ and remains nearly invariant with $t_{InAs}$. Increasing $t_{GaSb}$ results in a rise of $E_g$ whereas increasing $t_{InAs}$ leads to a lowering of only $CB_{min}$ and thereby reducing $E_g$.}
	\label{Eg}
\end{figure*}
\section{Results and Discussion}
\label{sec_result}
\subsection{k.p results}
\label{sec_kp}
The InAs/GaSb superlattice is distinguished by a broken band type-II alignment with non-common atomic arrangements which exhibits completely different properties compared to it's constituent materials. Therefore, the appropriate values of temperature dependent valence band offsets (VBO) are important to model such non-common atom structures as they are highly susceptible to the hetero-interface imperfections. It is already demonstrated in several earlier works that the interface region can ideally be of InSb or GaAs (or their higher order compounds) type based on the growth condition \cite{HauganPRB,razeghi_tb}, however, in most of the cases, InSb is the preferred choice as far as the accurate band modeling is concerned \cite{delmas2019comprehensive}. \\
\indent At first, we calculate the electronic band structure of a 8 ML/8 ML InAs/GaSb MWIR SL at a temperature of $T=77$ $K$ using the 8 band $\bf{k.p}$ method discussed earlier. The above configuration corresponds to the thicknesses of InAs ($t_{InAs}$) and GaSb ($t_{GaSb}$) layers to be roughly $2.4$ $nm$ each. In our simulation, we assume a fixed interface layer thickness ($t_{int}$) of 0.5 ML which is symmetric with respect to the junction. Therefore, a unit cell of the periodic structure as shown in Fig. \ref{uc} is formed with the arrangement \{InSb($t_{int}/2$)-InAs($t_{InAs}-t_{int}$)-InSb($t_{int}$)-GaSb($t_{GaSb}-t_{int}$)-InSb($t_{int}/2$)\} which corresponds to the period thickness of $d=t_{InAs}+t_{GaSb}$. The values of the parameters used in the $\bf{k.p}$ calculation are summarized in Table \ref{table1}. The calculated dispersion results along the in-plane ($E$-$k_{\parallel}$) and out-of-plane ($E$-$k_{\perp}$) directions are respectively plotted in Fig. \ref{Ek}(a) and \ref{Ek}(b) within the first BZ. The obtained value of bandgap ($E_g$) at 77 $K$ for the given configuration is 0.27 $eV$ which corresponds to a cut-off wavelength of 4.59 $\mu m$ that matches fairly well with the experimental bandgap values obtained in the range of 0.269-0.275 $eV$ \cite{kaspi_apl}. The conduction band (C1) electrons exhibit a strong dispersion along both the parallel and perpendicular directions which suggests a lower electron effective mass ($m_e^*$) and a high group velocity in either directions. On the other hand, the first heavy hole subband (HH1) remains almost dispersionless along the perpendicular direction and shows a steady dispersion in the parallel plane. This suggests that the heavy holes are strongly localized along the transport direction and conduct only against a strong field.\\
\begin{table}
	\centering
	\caption{ Material parameters of InAs, GaSb and InSb used for the $\bf{k.p}$ band structure calculation at temperature of 77K \cite{band_param,becer2019modeling,livneh2012k,delmas2019comprehensive}.}
	\begin{tabular}{|l|l|l|l|}
		\hline
		Parameters & InAs & GaSb & InSb \\
		\hline
		Lattice constant({\AA})& 6.0584& 6.0959 & 6.4794\\
		\hline
		Energy band gap at 0K($eV$)& 0.418 & 0.814 & 0.24\\
		\hline
		Varshini Parameter $\alpha $ $[meV/K$]& 0.276 & 0.417&0.320\\
		\hline
		Varshini Parameter $\beta$ [$K$]&93&140&70\\
		\hline
		Effective mass electron ($m_e^*$)& 0.022&0.0412& 0.0135\\
		\hline
		
		Luttinger parameter $\gamma1$  & 19.4& 11.84 & 32.4\\
		\hline
		Luttinger parameter $\gamma2$  & 8.545 & 4.25\ & 13.3\\
		\hline
		Luttinger parameter $\gamma3$  & 9.17 & 5.01 & 15.15\\
		\hline
		Interband mixing parameter Ep [$eV$] & 21.5 & 22.4 & 23.3\\
		\hline
		Spin orbit splitting (SO) [$eV$] & 0.38& 0.76 & 0.81\\
		\hline
		Valence band offset (VBO) [$eV$] & -0.56 & 0 & 0.03\\
		\hline
	\end{tabular}
	\label{table1}
\end{table}
\indent The wavelength range of the photo-response spectra of a material ideally depends on its bandgap and the band widths. In periodic infinite SL structures, the thicknesses of the constituent materials and the nature of interface play a vital role in determining the subband properties. Therefore, it is necessary to investigate the effect of varying $t_{InAs}$ and $t_{GaSb}$ on the bandgap and the relative band edge values. Figure \ref{Eg}(a) shows the variation of $E_g$ as a function of $t_{InAs}$ and $t_{GaSb}$ in a 2-D color plot. The conduction band minima ($CB_{min}$) and the heavy hole valence band maxima ($HH_{max}$) with respect to the zero energy reference are also plotted respectively in Figs. \ref{Eg}(b) and \ref{Eg}(c) as functions of the same parameters. Both $t_{InAs}$ and $t_{GaSb}$ are varied from 8 to 16 ML and the resultant bandgap spans in the range of 0.08-0.348 $eV$ which corresponds to the cut-off wavelength range of 3.5-15 $\mu m$ roughly covering the MWIR and LWIR range. One must notice that both $E_g$ and $CB_{min}$ exhibit similar patterns and are quite sensitive to the variations of both $t_{InAs}$ and $t_{GaSb}$. Since the InAs layers confines electrons and act as a potential well, an increase in $t_{InAs}$ results in a steady lowering of $CB_{min}$ towards the InAs conduction band edge. On the other hand, GaSb layers isolate the overlap of electrons between the adjacent InAs layers and act as a barrier for them. This results in a rise of $CB_{min}$ with increasing $t_{GaSb}$. In contrast to the above picture, $HH_{max}$ slowly moves closer to the valence band maxima of GaSb with the rise of $t_{GaSb}$ and remains nearly invariant with $t_{InAs}$ due to the strong localization of heavy holes in GaSb layers. This study encompasses a key aspect of band-engineering of T2SL structures for a wide infrared range and can be further utilized to design T2SL-based unipolar or bipolar barrier layers. \\
\begin{figure}[!htbp]
	\centering
	\subfigure[]{\includegraphics[height=0.225\textwidth,width=0.225\textwidth]{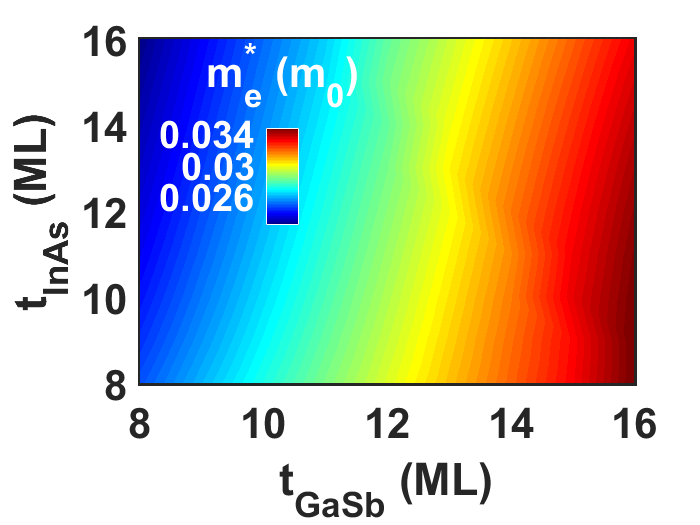}\label{ma}}
	\quad
	\subfigure[]{\includegraphics[height=0.225\textwidth,width=0.225\textwidth]{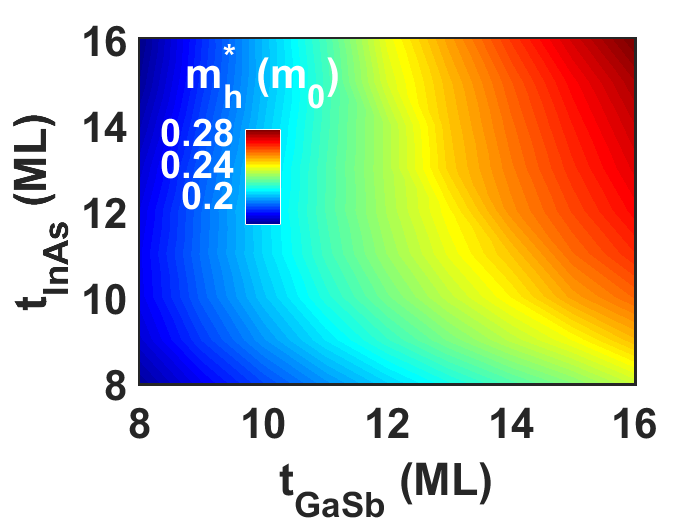}\label{mb}}
	\quad
	\caption{Electron and hole effective mass of T2SL as functions of layer thicknesses in 2D color plots: Effective mass of (a) electron ($m_e^*$) and (b) hole ($m_h^*$) are plotted as functions of $t_{InAs}$ and $t_{GaSb}$ (in ML). $m_e^*$ rises sharply with the increase of $t_{GaSb}$ and reduces marginally with increasing $t_{InAs}$. On the other hand, an increase in both $t_{InAs}$ and $t_{GaSb}$ leads to the rise of $m_h^*$.}
	\label{meh}
\end{figure}
The carrier transport properties in a superlattice structure in the form of miniband or hopping conduction are strongly governed by the quantum mechanical tunneling between the adjacent material layers. Since the tunneling probability has an exponential dependence on the carrier effective masses, the calculation of effective mass as a function of layer thicknesses holds much significance. In InAs/GaSb SL, effective masses are less influenced by the interaction between the valence and conduction band as they mainly depend on the electron (hole) wavefunction overlap between the adjacent InAs (GaSb) layers \cite{delmas2019comprehensive}. The parallel and perpendicular effective masses of electrons ($m_{e\parallel}^*$, $m_{e\perp}^*$) and holes ($m_{h\parallel}^*$, $m_{h\perp}^*$) are respectively calculated from the second derivatives of $E$-$k_{\parallel}$ and $E$-$k_{\perp}$ relations at the BZ center. The total effective mass is therefore calculated as $m_{e(h)}^*=\left( m_{e(h)\parallel}^*\right)^{2/3} \left( m_{e(h)\perp}^*\right)^{1/3}$. In Figs. \ref{meh}(a) and \ref{meh}(b), we respectively show the variations of $m_e^*$ and $m_h^*$ with respect to $t_{InAs}$ and $t_{GaSb}$. It can be seen that for the 8 ML/8 ML configuration $m_e^*=0.0245$ which is closer to that of the bulk InAs value and $m_h^*=0.16$, way lesser than both bulk InAs and GaSb. However, due to the strong confinement, $m_{h\perp}^*$ comes in the range of tens. One must notice that $m_e^*$ increases sharply with $t_{GaSb}$ due to the suppression of adjacent InAs layers carrier overlap and falls marginally with $t_{InAs}$, thereby maximizing at lower $t_{InAs}$ and higher $t_{GaSb}$ values. On the other hand, $m_h^*$ rises with both $t_{InAs}$ and $t_{GaSb}$ due to its identical values in both the materials. This study thus offers an engineering guideline in estimating the range of dark current, noise level and quantum efficiency while designing T2SL detectors.\\
\subsection{NEGF results}
\label{sec_negf}
\begin{figure}[!htbp]
	\centering
	\subfigure[]{\includegraphics[height=0.08\textwidth,width=0.4\textwidth]{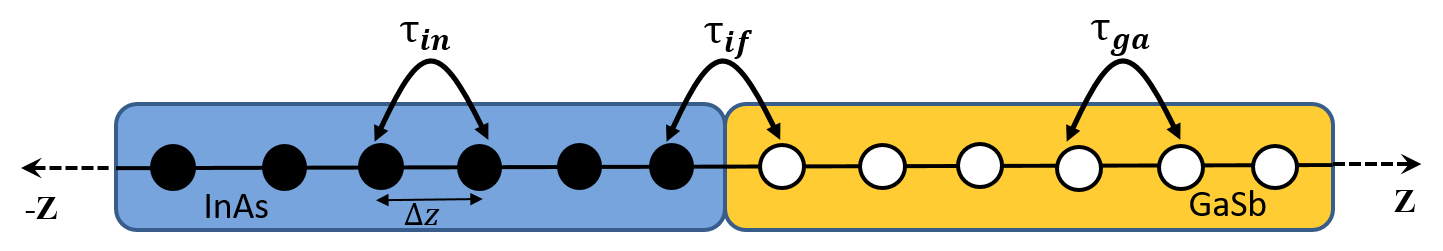}\label{ucna}}
	\quad
	\subfigure[]{\includegraphics[height=0.1\textwidth,width=0.4\textwidth]{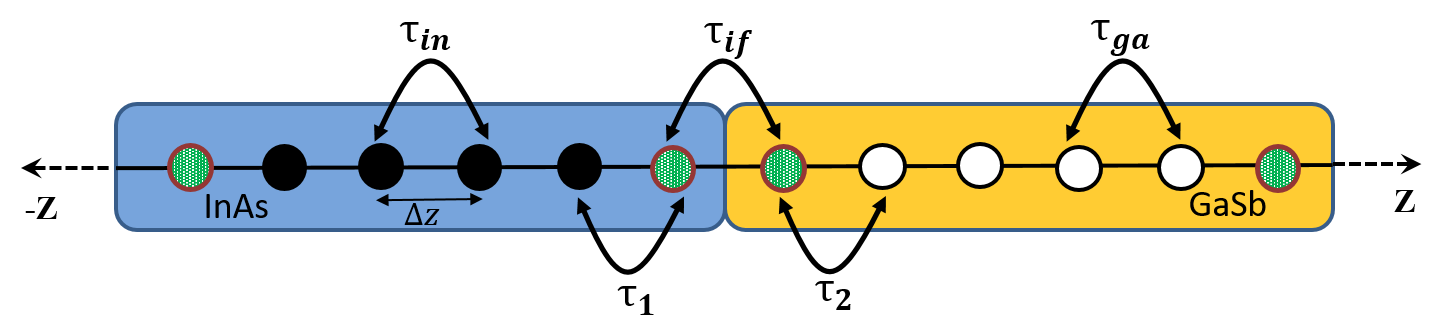}\label{ucnb}}
	\quad
	\caption{1D schematic of T2SL unit cell: Nearest-neighbour 1D atomic chain representation of a T2SL unit cell having (a) no-interface atom , and (b) common-interface atom. The green atoms, forming the interface are modeled with GaAs material parameters.}
	\label{ucn}
\end{figure}

\begin{figure*}[]
	\centering
	\subfigure[]{\includegraphics[height=0.25\textwidth,width=0.25\textwidth]{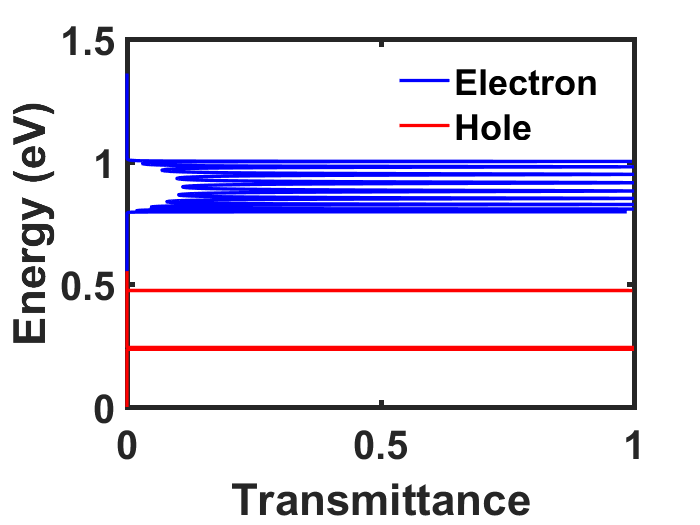}\label{minia}}
	\quad
	\subfigure[]{\includegraphics[height=0.25\textwidth,width=0.25\textwidth]{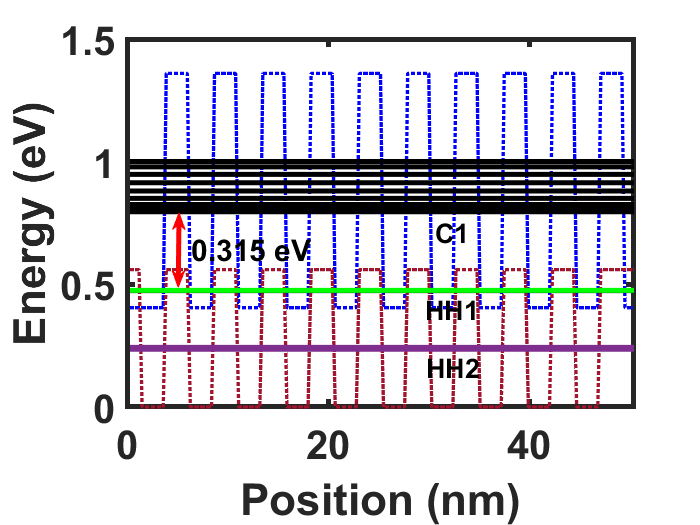}\label{minib}}
	\quad
	\subfigure[]{\includegraphics[height=0.25\textwidth,width=0.25\textwidth]{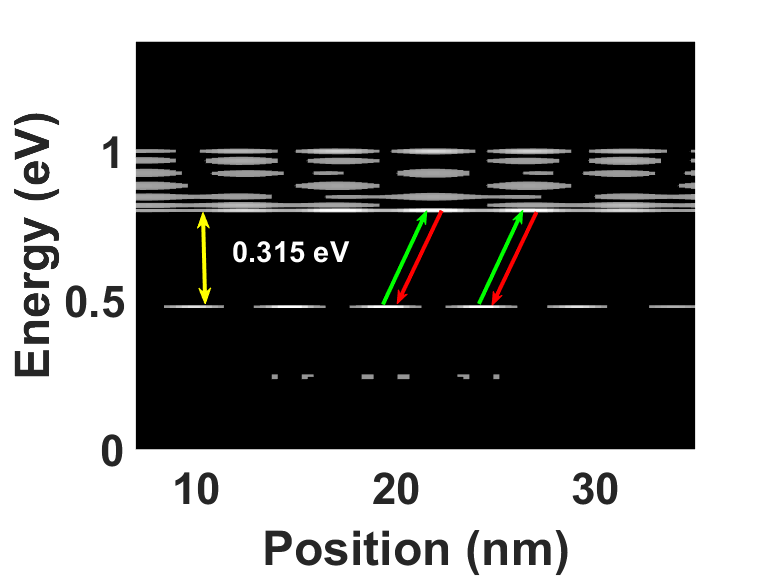}\label{minic}}
	\quad
	\caption{Transmission function and miniband formation: (a) Electron and hole transmission probabilities of a 10 period 8 ML/8 ML T2SL plotted with respect to energy. (b) Minibands derived from the transmission peaks are shown on the energy flat band diagram with respect to the position. (c) Local density of states (LDOS) plotted in a grey scale 2-D plot in the position and energy space indicates the spatial separation of electron and hole confinements and the possible formation of minibands.}
	\label{mini}
\end{figure*}

\begin{figure}[!htbp]
	\centering
	\subfigure[]{\includegraphics[height=0.225\textwidth,width=0.225\textwidth]{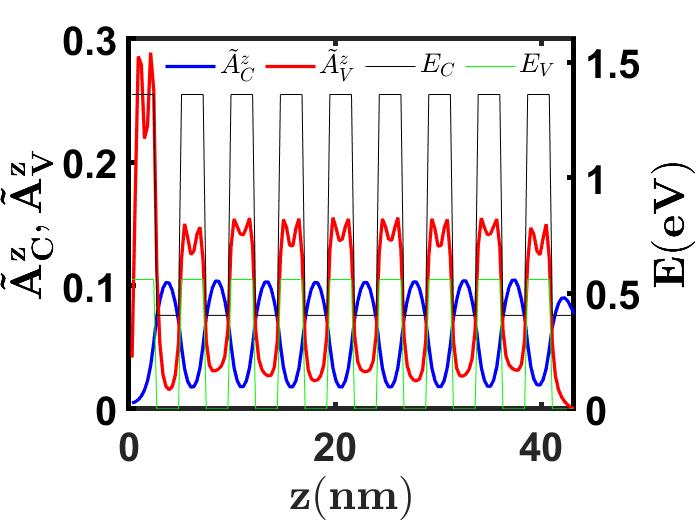}\label{overlapa}}
	\quad
	\subfigure[]{\includegraphics[height=0.225\textwidth,width=0.225\textwidth]{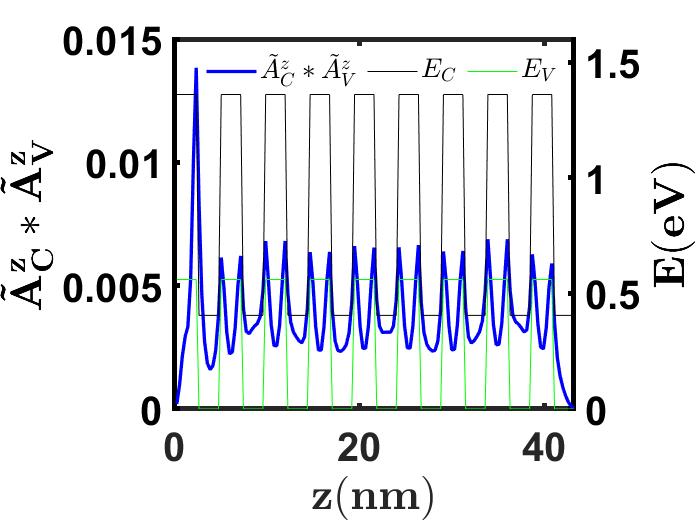}\label{overlapb}}
	\quad
	\caption{Spatial distribution of available states and their interband overlap: plot of the (a) number of available states in the conduction ($\Tilde{A}^z_C$) and heavy hole band ($\Tilde{A}^z_V$), and (b) their spatial product ($\Tilde{A}^z_C*\Tilde{A}^z_V$) with respect to the position ($z$). The overlap plot indicates the peak absorption right at the interface with a sharp falling trend deep inside the layers.}
	\label{overlap}
\end{figure}

\begin{figure}
    \centering
    \includegraphics[height=0.35\textwidth,width=0.4\textwidth]{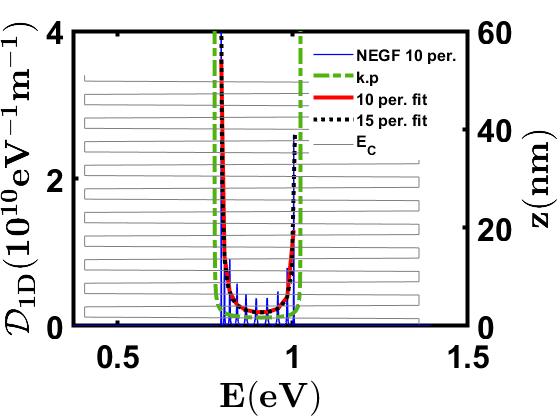}
    \caption{One dimensional density of states: 1D-DOS calculated from the $k.p$ theory for an infinite structure and from the NEGF method with different SL periods are shown for comparison. There is a good agreement between them apart from few disparity in the width and symmetry. The disparity minimizes with increasing SL periods.}
    \label{dos}
\end{figure}

\begin{figure}[!htbp]
	\centering
	\subfigure[]{\includegraphics[height=0.225\textwidth,width=0.225\textwidth]{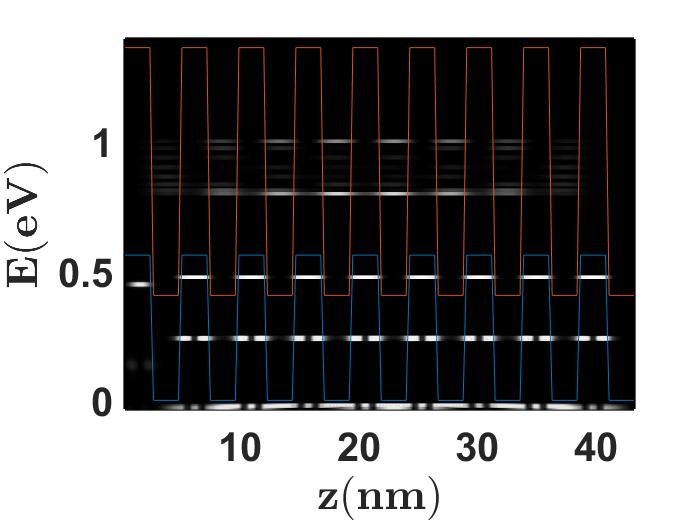}\label{diffldosa}}
	\quad
	\subfigure[]{\includegraphics[height=0.225\textwidth,width=0.225\textwidth]{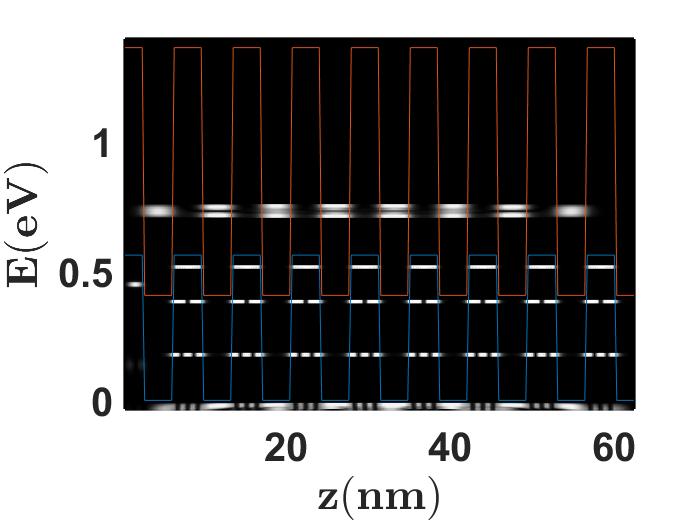}\label{diffldosb}}
	\quad
	\subfigure[]{\includegraphics[height=0.225\textwidth,width=0.225\textwidth]{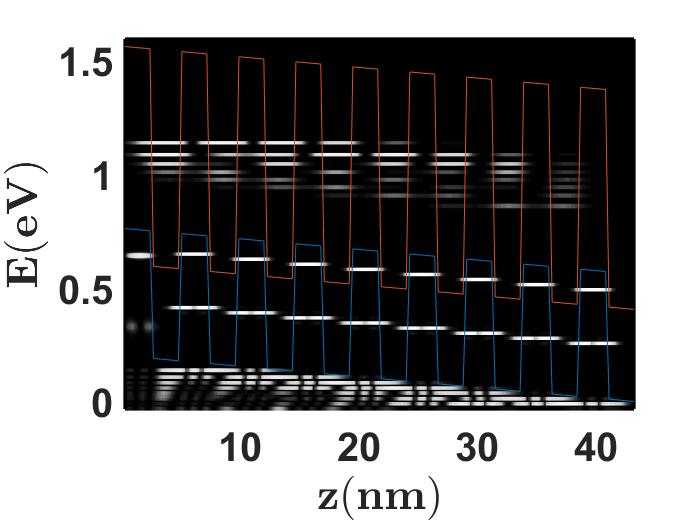}\label{diffldosc}}
	\quad
	\subfigure[]{\includegraphics[height=0.225\textwidth,width=0.2225\textwidth]{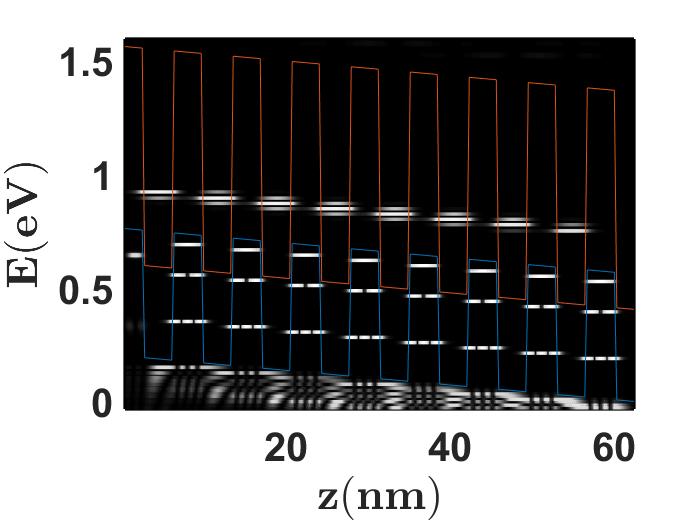}\label{diffldosd}}
	\quad
	\caption{LDOS profile of 8 ML/8 ML and 12 ML/12 ML T2SL are shown under flat band condition (a-b) and at a built-in potential of 0.2 $eV$ (c-d). Increasing layer thicknesses result in more number of bound states inside the band with a strong localization of the carriers in respective layers indicating an unlikely situation for miniband formation.}
	\label{diffldos}
\end{figure}

\begin{figure}[!htbp]
	\centering
	\subfigure[]{\includegraphics[height=0.225\textwidth,width=0.225\textwidth]{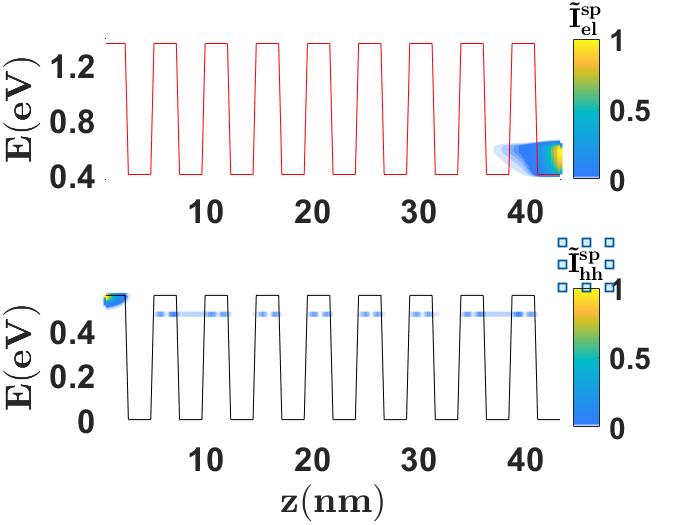}\label{Ia}}
	\quad
	\subfigure[]{\includegraphics[height=0.225\textwidth,width=0.225\textwidth]{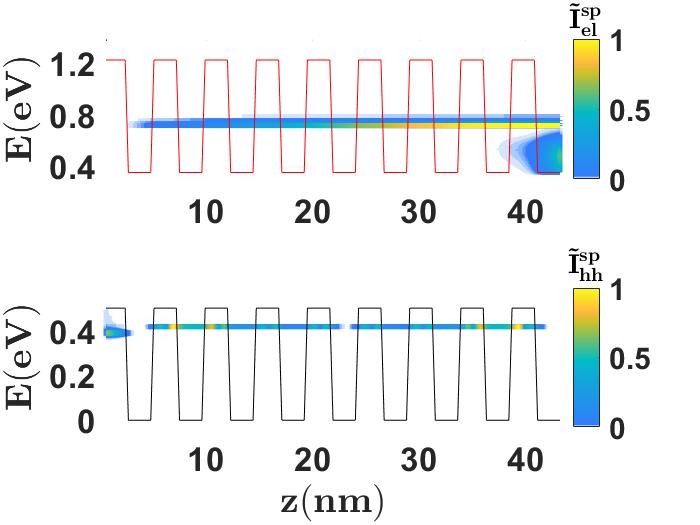}\label{Ib}}
	\quad
	\caption{Normalized spectral current of electron ($I^{sp}_{el}$) and heavy hole ($I^{sp}_{hh}$) are shown with respect to position and energy at (a) 77K and (b) 300K. Raising the temperature leads to an increasing population of the miniband states resulting in a significant current flowing through them.}
	\label{spectralI}
\end{figure}

\begin{figure}
    \centering
    \includegraphics[height=0.4\textwidth,width=0.4\textwidth]{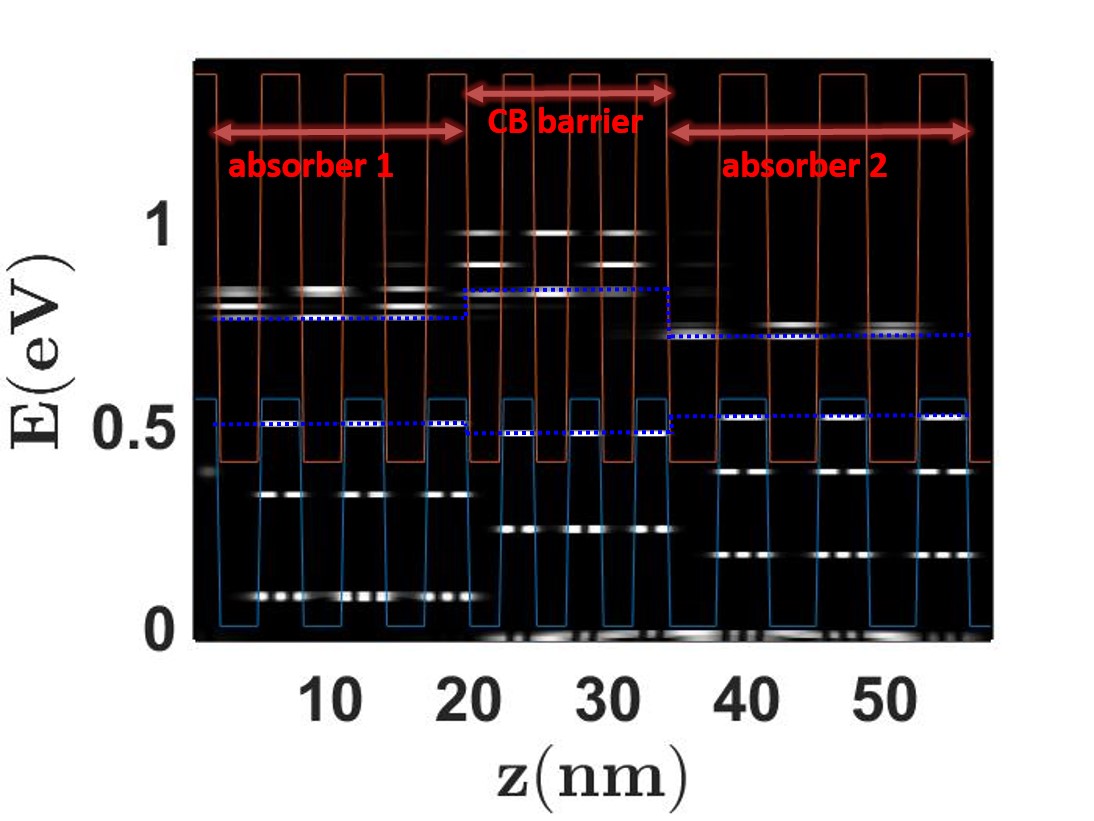}
    \caption{LDOS profile of a non-periodic T2SL structure having three different SL configurations of 3 periods each. The whole structure resembles a unipolar barrier dual band detector where the two absorber regions are modeled with 10 ML/10 ML and 12 ML/12 ML T2SL and the electron blocking barrier region in the middle is made of 8 ML/8 ML T2SL. The band diagram shown in blue dotted line clearly indicates the desired band alignment features.}
    \label{nBn}
\end{figure}

Having obtained the band structure, We now shift our attention towards examining the miniband, carrier localization and spectral current properties of a finite sized T2SL structure both in position and energy space using the NEGF approach. 
To model a T2SL structure using single-band NEGF, we first assume that the atoms of the constituting materials are arranged in an one-dimensional chain along the z-axis with nearest-neighbour interaction as shown in Fig. \ref{ucn}. The inter-atomic coupling parameters between two adjacent atoms are denoted as $\tau_{in}$ and $\tau_{ga}$ respectively for InAs and GaSb, and the coupling parameter between the two atoms on the opposite sides of the interface is given by $\tau_{if}$. These values are calculated from the parabolic dispersion relations of the individual materials in terms of their effective masses. The material parameters used in the NEGF simulation are summarized in Table \ref{negftable}. Based on the interfacial arrangement of atoms, we consider two different configuration schemes, namely, (i) No-interface atoms and (ii) Common-interface atoms, respectively shown in Figs. \ref{ucn}(a) and \ref{ucn}(b). In the following, we discuss both the configurations and present a comparative study of the simulated results.\\
\begin{table}
	\centering
	\caption{ Material parameters used in the NEGF simulation \cite{livneh2012k,band_param}.}
	\begin{tabular}{|l|l|l|}
		\hline
		Parameters & InAs & GaSb \\
		\hline
		Electron effective mass ($m_e^*$) & 0.023 & 0.041\\
		\hline
		Heavy hole effective mass ($m_h^*$) & 0.41 & 0.4\\
		\hline
		VBO [$eV$] @77K & -0.56 & 0\\
		\hline
		VBO [$eV$] @300K & -0.50 & 0\\
		\hline
	\end{tabular}
	\label{negftable}
\end{table}
\textbf{No-interface atoms}: In this scheme, the interface is assumed to be abrupt between the InAs and GaSb layers and there is no separate interfacial layer of atoms. For such an arrangement of a ten period 8 ML/8 ML superlattice, at first, we calculate the electron and hole transmission probabilities in the ballistic limit \cite{myTED,myPRA,myTF,pankaj}, given by Eq. \eqref{transeq} at 77K and plot them in Fig. \ref{minia}. The energy flat band diagram of the given structure is shown in Fig. \ref{minib} where the conduction and valence band edges of the two contact regions are intentionally terminated to their respective lowest and highest values to get perfect transmission peaks. The SL minibands formed in the conduction and valence bands as depicted in Fig. \ref{minib} are interpolated from the transmission peaks. One can notice from Figs. \ref{minia} and \ref{minib} that the electrons, owing to their lower effective mass, exhibit a sparse transmission spectra (C1) where the number of transmission peaks equals the number of SL periods considered which indicates the number of bound eigen states inside the band. The heavy holes, on the other hand, due to their large effective mass, form two sharp and dense transmission channels (HH1 and HH2) inside the valence band with the same number of allowed bound states. The LDOS, given by Eq. \eqref{LDOS}, of the aforementioned structure as shown in Fig. \ref{minic}, exhibits the localization profile of the carriers in C1 and HH1 bands which clearly indicates the spatial separation of the electron and hole confinements respectively in InAs and GaSb layers. The finite overlap between them determines the strength of absorption. The delocalization of conduction band electron wavefunctions between the neighbouring InAs layers also points towards a possible formation of spatially continuous miniband, however, the strongly localized heavy holes in the GaSb layers do not allow to form the same. The bandgap of the given structure is calculated from the difference between the bottom of C1 and top of HH1 which in this case comes out to be 0.315 $eV$. It is quite evident that the overestimation of the bandgap happens due to the absence of an interfacial layer, in compliance with the $\bf{k.p}$ calculation. The correct value of bandgap can be obtained through an interface modeling and will be shown later using our common-interface model. However, one can gain deeper physical insights into the spatial as well as spectral properties of the individual bands using this simple model as elaborated in the following.\\
\indent To investigate the spatial confinement and interband overlap properties of the carriers, we follow a more realistic simulation approach by taking into account the effects of dephasing through the incorporation of phase and momentum-breaking elastic scattering processes. These processes are inherently present in the system even at a low temperature and can be generally attributed to the interaction between the electrons and low-energy acoustic phonons or the surface/interface scattering mechanisms. Such dephasing phenomenon are incorporated through scattering self-energies ($\Sigma_S$) analogous to B\"uttiker-probes \cite{DattaQT} which necessitates a self-consistent solution approach in the NEGF framework. In setting up the simulation, we further consider that the left and right contacts are respectively of GaSb and InAs type which makes them carrier-selective in terms of hole and electron injection. With these considerations, we simulate an eight period 8 ML/8 ML T2SL structure and plot the total number of states of the conduction ($\Tilde{A}^z_C$) and heavy hole band ($\Tilde{A}^z_V$), and their spatial product ($\Tilde{A}^z_C*\Tilde{A}^z_V$) with respect to the position ($z$) respectively in Fig. \ref{overlapa} and Fig. \ref{overlapb}. $\Tilde{A}^z_{C/V}$ is calculated by integrating the LDOS ($A_{C(V)}(z,z,E)$) over energy and is given by $\Tilde{A}^z_{C(V)}(z)=\int A_{C(V)}(z,z,E)dE$.\\
\indent Figure \ref{overlapa} clearly shows that the electron and hole states are uniformly distributed throughout the lattice respectively in the InAs and GaSb layers and while the electronic confinement peaks in the middle of InAs layers, the heavy hole states exhibit multiple peaks in the GaSb layers. This happens due to the contribution from multiple heavy hole subbands (HH1 and HH2 in this case) to the $\Tilde{A}^z_V$. To have a better understanding of the microscopic picture, one must note that unlike the other methods \cite{Kaya_miniband,becer2019modeling}, the wavefunctions do not peak at the center of the lattice, rather they spread evenly over the entire structure owing to the re-distribution of allowed states arising due to the dephasing effects. The beauty of this method lies in the practical consideration of multiple momentum and phase-breaking processes running inside the system which tend to smear out the local density of states throughout the lattice and thereby manifesting a much more realistic picture of the carrier localization and band properties. The strength of absorption, on the other hand, being primarily governed by the amount of interband overlap of the electron and hole states, becomes a strong function of the layer thicknesses. Therefore, the spatial product of $\Tilde{A}^z_C$ and $\Tilde{A}^z_V$ as plotted with respect to position in Fig. \ref{overlapb} plays a major role in the present context. One notices that the overlap peaks right at the interfaces and falls sharply away from it. This suggests that the maximum amount of absorption takes place at the junctions of the material layers and one should have thin layer T2SL structures to maximize the absorption. However, this enhances the possibility of recombination too which can severely degrade the detector performance. This study, therefore, provides significant information on the design strategies to achieve an optimum trade-off between the absorption and recombination factors.\\
\indent In quantum confined heterostructures, the knowledge of 1-D DOS ($\mathcal{D}_{1D}$) becomes indispensable in determining the carrier concentrations and their vertical transport properties. This quantity can be calculated from the out-of-plane dispersion ($\epsilon_n(q)$) relation obtained from the $\bf{k.p}$ results with periodic boundary conditions and is given by $\mathcal{D}_{1D}(E)=\frac{2}{\pi}\{\frac{\delta \epsilon_n(q)}{\delta q}\}^{-1}_{q(\epsilon_n=E)}$ \cite{Aeberhard_apl}. Using this relation, in Fig. \ref{dos}, we plot the conduction band 1D-DOS, calculated from the dispersion profile depicted in Fig. \ref{Ekb}, as a function of energy. The sharp features at the two ends arising from the flat dispersion at the corners and the center of the first BZ, indicate the two edges of the miniband which also provides an estimate of the bandwidth. We further show that using only a finite structure in the NEGF approach, the 1D-DOS at zero transverse momentum can also be evaluated by integrating the LDOS over the entire finite SL structure, given by $\mathcal{D}_{1D}(E)=\frac{1}{L_z\pi}\int dz A(z,z,E,k_{\parallel}=0)$, where $L_z$ denotes the length of the structure. For comparison, we display the 1D-DOS profiles obtained from NEGF for different number of SL periods in Fig. \ref{dos} alongside the $\bf{k.p}$ result. The plot shows a fair amount of agreement between the 1D-DOS derived from the finite and infinite structure in terms of the bandwidth and shape except for some dissimilarities in the magnitude and symmetry. However, one must notice that as we increase the number of periods, these discrepancies start to disappear, resulting in a more accurate 1D-DOS profile inside the band. \\
\indent In order to understand the physical properties of carrier localization and the miniband formation, and their dependency on the layer thicknesses, we display the LDOS profile of two different SL configurations, 8 ML/8 ML and 12 ML/12 ML, respectively in Fig. \ref{diffldosa} and Fig. \ref{diffldosb} under flat-band conditions and in Fig. \ref{diffldosc} and Fig. \ref{diffldosd} under a finite built-in potential of 0.2 $eV$. These figures collectively depict some interesting signatures of the T2SL band properties. As the layer width increases, the miniband widths converge at a rapid rate along with the reduction of bandgap and more number of subbands appear inside the band. The LDOS of the 8 ML/8 ML structure, under both the flat-band and built-in conditions, reveals the strong delocalization of the electronic states and the formation of the conduction miniband. However, with the increasing layer thicknesses, these states start to localize more and more in the InAs layers resulting in a weaker coupling of periods which ceases the possibility of miniband formation as visible from the LDOS of the 12 ML/12 ML structure. Specifically, the thickness of the electron blocking GaSb layers play the key role here in tailoring the strength of coupling and the amount of delocalization. On the other hand, the heavy hole states in all these cases remain strongly localized in the GaSb layers which suggests that the vertical transport of heavy holes in the MWIR range of operation is mostly governed by the tunneling conduction. This study, therefore, helps to find out the critical thicknesses of the layers that distinguishes the mode of carrier transport and determines the spectral properties.\\
\indent A more detailed picture of the carrier transport through the miniband states inside the band can be obtained from the spectral current properties of electrons and holes over the entire structure. Within elastic scattering, the current flowing between any two adjacent atoms at a particular energy, given by Eq. \eqref{Ieqn}, is plotted for the 8 ML/8 ML structure at two different operating temperatures of 77K and 300K, respectively in Fig. \ref{Ia} and Fig. \ref{Ib}. For a qualitative discussion, the spectral currents ($I^{sp}$) are normalized ($\tilde{I}^{sp}$) with respect to their maximum values. The magnitude of $I^{sp}_{el(hh)}$ and the coupling between the periods largely depend on the electron (heavy hole) population in the conduction (valence) miniband as expressed by the correlation functions $G^{n(p)}$. These functions, as given by Eq. \eqref{GnGp}, are directly related to the equilibrium Fermi functions of the contacts, characterized by their electrochemical potential ($\mu$) and temperature. In this simulation, we set $\mu$ of both the contacts at 0.2 $eV$ above the conduction band edge of InAs. At 77K, the conduction miniband states are hardly populated with electrons and thereby showing negligible conduction through it. Conduction only takes place around the $\mu$ of the right InAs contact. However, the heavy hole states, although weakly populated, exhibit a signature of finite and spatially discontinuous current profile through the left GaSb contact. On the other hand, at a higher temperature of 300K, both these states are largely populated with carriers and conduct significantly with a broader spectral profile of current. This can be easily verified from Fig. \ref{Ib} where the electrons and heavy holes exhibit a clear signature of miniband transport, although the coupling of periods is still weak in the heavy hole band. Under finite built-in potential, the current spectrum can become much more broadened with the inclusion of inelastic scattering events between the electrons and optical phonons.\\
\indent The applicability of this method is further extended towards a comprehensive understanding of the barrier-based complex detector structures. The family of next-generation high performance detectors is characterized by their high temperature operation and multi-color detection. This necessitates the integration of multiple absorber regions along with the inclusion of unipolar or bipolar carrier blocking barrier layers in a single device. In T2SL based structures, all these layers can be realized by varying the SL configurations using different shutter sequences in the growth chamber. In order to understand the different band alignments in such complex structures, one needs to separately evaluate the band properties of each SL region assuming perfect periodicity and combine them together for comparison. However, the NEGF method in the dephasing picture, reveals its supremacy in obtaining the band profiles of such complex structures in an integrated simulation approach even when the SL periodicity is lifted. In doing so, we consider a T2SL structure that contains two absorber regions and one unipolar electron blocking barrier layer. The two absorber regions are configured using a 10 ML/10 ML and 12 ML/12 ML T2SL and the barrier region is configured using a 8 ML/8 ML structure. The LDOS profile of the given structure is displayed in Fig. \ref{nBn}. For the sake of computational complexity, we have only considered 3 periods for each of these layers while ideally the number can be stretched further. The composite band diagram of this structure is shown in a blue dotted line which is drawn on the bottom of the conduction miniband and the top of the first heavy hole band. The two absorber regions clearly reveal different bandgaps which indicates dual-color detection under different biasing condition. The barrier layer, on the other hand, shows a higher value of bandgap and a large conduction band offset while depicting a negligible valence band offset in compliance with the requirement of unipolar electron blocking layer. The minute differences between the heavy hole states in different regions both in the energy and position space can also be clearly observed. This study is thus extremely fruitful for a qualitative understanding of the band profiles and localization in any complex SL structures with broken-periodicity which eventually justifies the effectiveness of the method adopted.\\
\textbf{Common-interface atoms}: The study, so far, was aimed to build adequate understanding of the band and transport properties in a qualitative manner. However, the bandgap, being overestimated earlier, should be predicted accurately for a given T2SL structure in order to get the right absorption wavelength. From our earlier discussion, we understand that the appropriate modeling of the interface can give rise to correct bandgap values. Several earlier works were dedicated to achieve bandgap values within the acceptable limit of errors through interface engineering \cite{delmas2019comprehensive,Thi_2019}. This work, too, adopts a simple interface modeling approach by inserting different atoms common to either side of the interface as shown in Fig. \ref{ucnb}. Unlike the $\bf{k.p}$ method, we consider that the interface atoms only change the inter-atomic coupling parameters and do not form a separate material layer. In our simulation, we assume that the interface atoms are of GaAs type and the coupling parameters around the interface are accordingly calculated from GaAs effective masses ($m_e^*=0.067m_0$, $m_h^*=0.4m_0$). For the earlier 8 ML/8 ML structure at 77K, the bandgap using this model is calculated as 0.278 $eV$ which matches reasonably well with the $k.p$ results and the experimental data. However, due to the large electron effective mass of GaAs, the conduction miniband width reduces substantially. A comparative study of the bandgap ($E_g$) and the conduction miniband width $\Delta_{C1}$ of the 8 ML/8 ML structure obtained from both these methods and experiments is presented in Table \ref{8ML8ML}. To validate this model for bandgap estimation, we further simulate other structures and compare them in Table \ref{compareEg} with the $k.p$ and the available experimental data in the literature. While the bandgap matches well in the common-interface atom model, the no-interface model fairly estimates the bandwidth which improves further with the increasing number of SL periods. In future outlook, we believe that a rigorous multi-band tight-binding modeling in the NEGF formalism can lead to even more accurate results in terms or band and transport characteristics.
\begin{table}[ht]
\caption{Comparison of calculated and experimental \cite{kaspi_apl} bandgap ($E_g$) and conduction band width ($\Delta_{C1}$) of a 8 ML/8 ML InAs/GaSb T2SL at 77K.}
    \centering
    \begin{tabular}{|c|c|c|c|c|}
    \hline
        8 ML/8 ML & \multicolumn{2}{c|}{NEGF} & k.p & Experimental\\
        \cline{2-3}
        InAs/GaSb @77K & No-IF & GaAs-IF & InSb IF & Result\\
        \hline
        Bandgap ($E_g$) eV & 0.315 & \textbf{0.278} & 0.27 & 0.269-0.277\\
        \hline
        CB width ($\Delta_{C1}$) eV & \textbf{0.21} & 0.13 & 0.235 & 0.22\\
        \hline
    \end{tabular}
    
    \label{8ML8ML}
\end{table}
\begin{table}[ht]
    \caption{Comparative study of bandgap ($E_g$) in $eV$ at 77K obtained from $k.p$ and NEGF methods with the available experimental data \cite{kaspi_apl,klein_exptdata} for different T2SL configurations.}
    \centering
    \begin{tabular}{|c|c|c|c|c|}
    \hline
        InAs/GaSb  & k.p & NEGF & NEGF & Experimental\\
        configurations &  & without-IF & with-IF & result\\
        \hline
        7 ML/8 ML & 0.305 & 0.36 & 0.31 & 0.3 \\
        \hline
        8 ML/8 ML & 0.27 & 0.315 & 0.278 & 0.269-0.277 \\
        \hline
        10 ML/10 ML & 0.225 & 0.25 & 0.215 & 0.22-0.23 \\
        \hline
        8 ML/12 ML & 0.315 & 0.33 & 0.285 & 0.304-0.308 \\
        \hline
    \end{tabular}
    \label{compareEg}
\end{table}

\section{Conclusion}
\label{conclu}
In conclusion, we have thoroughly investigated the carrier localization, miniband, and spectral transport properties of a InAs/GaSb based T2SL detector using NEGF based quantum transport approach. We have evaluated the band structure of an infinite T2SL structure with periodic boundary condition using the EFA based $\bf{k.p}$ technique and demonstrated the tunability of bandgap and DOS effective mass with respect to the changes in the constituent material layer thicknesses. Subsequently, using the NEGF model coupled to the elastic dephasing processes, we have calculated the LDOS of a finite sized periodic T2SL and show that the there is a close agreement in terms of the DOS and band edge positions between the finite and infinite structure. A systematic study on the carrier localization and spectral current has also been carrier out in order to understand the roles played by the design parameters such as layer thickness and SL period, temperature and built-in potential. We have further shown that a simple interface engineering in the finite structure can produce a fair estimation of the bandgap. Finally, the versatility of this model is depicted via the prediction of band alignment features of non-periodic barrier based multi-band T2SL structures with varying SL configurations which paves the way for efficient designing of next-generation detectors. \\
\section*{Acknowledgments}
The authors are grateful to Arup Banerji and Naresh from the Indian Space Research Organization (ISRO) for useful discussions. The authors acknowledge funding from ISRO under the ISRO-IIT Bombay Space Technology Cell. This work is also an outcome of the Research and Development work undertaken in the project under the Visvesvaraya PhD Scheme of Ministry of Electronics and Information Technology, Government of India, being implemented by Digital India Corporation (formerly Media Lab Asia). This work is also partially supported by the Science and Engineering Research Board (SERB), Government of India, Grant No EMR/2017/002853 and Grant No. STR/2019/000030, the Ministry of Human Resource Development (MHRD), Government of India, Grant No. STARS/APR2019/NS/226/FS under the STARS scheme.
\bibliography{reference}
\end{document}